\documentstyle[12pt]{article}
\begin{document}
\begin{titlepage}
\title{The Ladder Approximation in $QED_2$}
\author{
C. D. Fosco
 and
T. Matsuyama
\footnotemark[1]\\
Department of Physics, Theoretical Physics \\
University of Oxford \\
1 Keble Road, Oxford OX1 3NP, U.K.}
\date{}
\footnotetext[1]{Permanent address: Department of Physics, Nara University of
 Education, Takabatake-cho, Nara 630, Japan}
\vspace{1cm}
\begin{abstract}
We study the Schwinger-Dyson equation for the fermion self-energy in
massless and massive $QED_2$, in the ladder approximation.
When the fermion is massless (and the photon massless or massive),
we check the reliability of this approximation by comparing its solutions
with the exact ones.  They agree only when the photon is massless.
For a massive fermion and  massless photon, we show that there is no
consistent solution at all, the infrared divergences introduced by
the approximation forbidding even the trivial solution.
When both fermion and photon are massive, we find a non-perturbative
(extra) fermion mass generation (which survives in the limit when the
bare mass of the fermion tends to zero). We argue that, in this case,
the ladder approximation will provide reliable results if the bare masses
of the fields are large compared with the (dimensionful) coupling constant.
\vskip 3cm

Key words: Schwinger-Dyson, $QED_2$

$\phantom{Key words:}$

\vskip 3cm
\end{abstract}
\maketitle
\end{titlepage}
\baselineskip=21.5pt
\parskip=3pt
\section{Introduction}
The Schwinger-Dyson Equations~\cite{itz} (SDE) of a model encode
its perturbative and non-perturbative dynamics. Under some suitable
approximation schemes, it
is sometimes possible to reduce them to a more manageable set of
self-consistent equations involving the Green's functions of interest
for the phenomenon under consideration. This approach has been
extensively used for the study of non-perturbative effects in theories
involving fermions and vector fields, with particular interest in the
calculation of the fermionic self-energy, because of its relevance for
(non-perturbative) chiral symmetry breaking and mass
generation~\cite{pen}.

The simplest approximation for the calculation of the fermion self-energy
is the so-called ladder approximation. It
corresponds to summing over all the self-energy diagrams without
fermion loops and without crossing between the different internal photon
propagators.
Although this is the simplest approximation, one can  seldom
extract any exact analytical result in this framework\footnote{Considerable
effort has been spent in applying this approximation to $QED_4$.
Maskawa and Nakajima~\cite{mas} have proved that it predicts a
self-consistent solution with a massive fermion.
Fukuda and Kugo~\cite{fuk} converted the self-consistent integral
equation to a differential equation plus boundary conditions, simplifying
the analysis.}.

The different approximations used to simplify the SDE (large N, ladder) are 
not
always easy to justify quantitatively, because of the lack of other
calculations of the same objects by independent methods (except when
some reliable lattice computation is available). There are reasons
to suspect that in low dimensions the popular truncations of the
SDE could be in jeopardy. Indeed, the ladder approximation is
essentially
the Hartree-Fock method~\cite{fet}, which gives reliable results only when
the fluctuations (i.e., current-current correlations) are negligible,
and this could hardly be true in
low dimensions. In particular, this approximation could predict
a phase transition where there is none, as it happens when the mean-field
approximation is used in the one-dimensional Ising model, for example.
On the other hand, for some two-dimensional models one can solve
the SDE exactly~\cite{sta}, so the possibility presents itself of comparing
those
solutions with the ones given by the ladder approximation.

In this paper, we study analytically and numerically the SDE for the
self-energy of the fermion in $QED_2$ in the ladder approximation,
exploring four different situations regarding the bare masses of the
fields:
\begin{description}
\item[1)] Both fields massless (Schwinger Model),
\item[2)] Massless fermion and massive photon,
\item[3)] Massive fermion and massless photon,
\item[4)] Both fields massive .
\end{description}

For the cases 1) and 2), we take advantage of the fact
that the models are exactly solvable,
which enables us to give a precise test of the approximation. We
find that for 1) the approximation predicts the fermion mass to
be zero, which is consistent with the exact solution~\cite{sta}.
For 2) the numerical solution of the approximation predicts a
non-zero mass, which is wrong
\footnote{Note that in the Schwinger Model there is chiral symmetry breaking,
which manifests itself by a non-zero value of the chiral condensate
$ \langle {\bar \psi} \psi \rangle $,
but the fermion remains massless.} (see Appendix).

In the case 3), we  show that there is no consistent
solution at all, due to the wild infrared behaviour of the photon
propagator.

For  4), we present the numerical solutions of the corresponding
equation, using the iteration method. We show that, when  the
bare masses of both fields are large compared with the coupling
constant, the result provided by the ladder approximation is
reliable.

The structure of the papers is as follows: in section 2 we introduce
the model and obtain its corresponding ladder SDE for the fermion
propagator. In section 3 we present our analysis of the four different
cases, showing the corresponding numerical solutions, when they are
non-trivial. In the Appendix we present the exact solution of cases 1) and 
2).

\section{The model and its SDE}
All the cases we will consider can be obtained from the general Lagrangian
\begin{equation}
{\cal L \mit} \,=\, -\frac{1}{4} F_{\mu \nu}F^{\mu \nu} + \frac{1}{2} \mu^2
A_{\mu}A^{\mu} - \frac{1}{2} \lambda (\partial_\mu A^\mu)^2 + \bar{\psi}
(i \not \! {\partial} - e \not \!\!{A} - m ) \psi  ,
\label{1}
\end{equation}
where $\mu$ and $m$ are the photon and fermion masses, respectively.
$e$ is the coupling constant, which has dimensions of mass.
$\lambda$ is a covariant gauge fixing parameter when $\mu=0$. When
$\mu \neq 0$, it improves the ultraviolet
behaviour of the photon propagator without changing the physical
quantities (the vector current is conserved quantum mechanically).
It also allows us to take the massless photon limit safely.

>From eq.(\ref{1}), we obtain the free fermion propagator
\begin{equation}
S_0 (p) \,=\, \frac{i}{ \not \! p  - m }\;\;,
\label{2}
\end{equation}
and the free photon propagator
\begin{equation}
G^{\mu \nu}(k)\,=\, -i\,(\frac{g^{\mu \nu}-k^\mu k^\nu /\mu^2}{k^2-\mu^2} +
\frac{k^\mu k^\nu /\mu^2}{k^2-\mu_\lambda^2}) \;,
\label{3}
\end{equation}
where we have defined $\mu_\lambda^2 = \frac{\mu^2}{\lambda}$.

The SDE for the fermion self-energy is
\begin{equation}
S^{-1} (p) \,=\, S_0^{-1} (p) - \Sigma (p) \; ,
\label{4}
\end{equation}
where $S$ is the full fermion propagator, which we parametrize as
\begin{equation}
S(p) \,=\, \frac{i}{ A(p) \not \! p - B(p) }\;
\end{equation}
and $\Sigma$ is the (1PI) self-energy.
On the other hand, $\Sigma$ is given by
\begin{equation}
\Sigma(p) \,=\, (-ie)^2 \int \frac{d^2 k}{(2 \pi)^2} \, \gamma^\mu \,
S(k) \,\Gamma^\nu (p, k) \,
G'_{\mu \nu} (p-k) \; ,
\label{6}
\end{equation}
where $\Gamma^\nu$ is the full vertex function and $G'_{\mu \nu}$ is the full
photon propagator.
In the ladder approximation,  $\Gamma^\nu$ is replaced by the tree vertex
$\gamma^\nu$ and $G'_{\mu \nu}$ by the free propagator (\ref{3}).
Thus eq.(\ref{6}) is approximated by
\begin{equation}
\Sigma(p) \,=\, (-ie)^2 \int \frac{d^2 k}{(2 \pi)^2} \, \gamma^\mu \, S(k)
\,\gamma^\nu \,
G_{\mu \nu} (p-k) \; .
\label{7}
\end{equation}
The set of eqs.(5) and (\ref{7}), when expanded in powers of
$e$, corresponds to the series of `rainbow' Feynman diagrams.
They are equivalent to a set of two coupled integral equations for the scalar
functions $A$ and $B$:
\begin{equation}
B\,=\, m + \frac{e^2}{2\pi}\frac{\lambda + 1}{\lambda} \int^{\infty}_0 dk
\frac{kB}{A^2 k^2 + B^2}
\frac{1}{\sqrt{(p^2+k^2+\mu^2)^2-4p^2 k^2}}
\label{8}
\end{equation}
and
\begin{eqnarray}
A-1 & = & - \frac{e^2}{4 \pi} \frac{1}{p^2 \mu^2} \int^{\infty}_{0} dk
\frac{kA}{A^2 k^2 + B^2}
( \frac{(p^2+k^2)^2 + \mu^2 (p^2+k^2)-4p^2 k^2}{\sqrt{(p^2+k^2+\mu^2)^2-
4p^2 k^2}} \nonumber \\
& - & \frac{1}{\lambda} \frac{(p^2+k^2)^2 + \mu_\lambda^2 (p^2+k^2)-
4p^2 k^2} {\sqrt{(p^2+k^2+\mu_\lambda^2)^2-4p^2 k^2}})
\label{9}
\end{eqnarray}
(after Wick rotation and angular integration).
Note that in the Feynman gauge ($\lambda=1$) eq. (\ref{9}) becomes just $A=1$
and eq.(\ref{8}) yields
\begin{equation}
B(p^2)\,=\,m + \frac{e^2}{\pi} \int^{\infty}_0 dk \frac{k B(k^2)}{k^2 +
{B(k^2)}^2}
\frac{1}{\sqrt{(p^2+k^2+\mu^2)^2-4p^2 k^2}} \; ,
\label{10}
\end{equation}
which is the only equation we have to consider. For latter convenience, we
rewrite it in two different ways.  First, defining $x = p^2$ and $y = k^2$,
eq.(\ref{10}) becomes
\begin{equation}
B (x) \,=\, m + \frac{e^2}{2 \pi} \int dx \, \frac{1}{\sqrt{
(x + y + \mu^2)^2 - 4 x y}} \,\frac{B(y)}{y + B^2 (k^2) }\;.
\label{11}
\end{equation}
On the other hand, for $\mu \neq 0$, we can simplify the treatment further
by working in terms of a dimensionless function
$b$ such that $B (p^2) = \mu \, b(p^2 / \mu^2)$.
In terms of $b$, eq.(\ref{11}) becomes
\begin{equation}
b(s) \,=\, \frac{m}{\mu} + \frac{e^2}{2 \pi \mu^2} \,
\int_0^{\infty} dt \frac{1}{\sqrt{ (s + t + 1)^2 - 4 s t}}
\,\frac{b(t)}{t + b^2 (t) }\;.
\label{12}
\end{equation}
Now we study eqs.(\ref{11}) and (\ref{12}) for
the cases 1)-4).

\section{Self-consistent solutions}
\subsection{The case $m=0$ and $\mu =0$ (Schwinger Model)}
In this case eq.(\ref{11}) reduces to
\begin{equation}
B (x) \,=\,\frac{e^2}{2 \pi} \int dy \, \frac{1}{\mid x - y \mid}
\,\frac{B(y)}{y + B^2(y) }\;.
\label{13}
\end{equation}
Although the kernel is singular at $y=x$, it might still be possible to
get a finite answer for the rhs of eq.(\ref{13}). Let us see that this
is not he case.

Assume that $B(x_0)$ is finite and non-zero in a neighbourhood of at least
one point $x_0$, then the lhs of eq.(\ref{13}) at $x = x_0$ is, of
course, finite and non-zero.
But evaluating the rhs of (\ref{13}) for $x = x_0$
we get an infinite result, thus implying that $B(x_0)$ should
be infinite, contradicting the original assumption.
We conclude that the only solution of (\ref{11}) is the trivial one:
$B=0$\footnote{ An infinite $B$ is not a solution, because it produces a
$0$ on the rhs of (\ref{11}), and $\infty$ on the lhs.}.

\subsection{The case $m=0$ and $\mu \neq 0$}
As $\mu \neq 0$, we work in terms of the dimensionless function
$b$ of eq.(\ref{12}), which reduces to
\begin{equation}
b(s) \,=\, \frac{e^2}{2 \pi \mu^2} \,
\int_0^{\infty} dt \frac{1}{\sqrt{ (s + t + 1)^2 - 4 s t}}
\,\frac{b(t)}{t + b^2 (t)}\;.
\label{14}
\end{equation}
We could not find any analytical solution of eq.(\ref{14}), however,
assuming that the solutions of eq.(\ref{14}) are decreasing functions,
the following inequalities hold:
\begin{equation}
\frac{b(t)}{t + b^2 (t)} \leq \frac{1}{\sqrt{t + b^2
(t)}} \leq \frac{1}{\sqrt{t}} \;,
\label{15}
\end{equation}
which inserted into eq.(\ref{14}) yield a constraint for $b(0)$
\begin{equation}
0 \, \leq \, b(0) \, \leq \, \frac{e^2}{2 \mu^2} \;.
\label{16}
\end{equation}
We have solved eq.$(\ref{14})$ numerically by the usual iteration
method, finding that there  are
non-trivial solutions for any (non-zero) value of the parameters.
They fall in the range predicted by (\ref{16}). In Figure
1 we show the behaviour of the self-consistent solution for
different values of the parameters
wich appear in the form of a dimensionless ratio $g = \frac{e^2}{2
\pi \mu^2}$. In terms of $g$, (\ref{16}) reads: $0 \,\leq \, b(0) \leq
\, \pi g$.

In Figure 2, we plot the value of the self-energy at zero-momentum,
as a function of the dimensionless coupling $g$. The numerical data
fit very well with a power-law behaviour: $b(0) \propto g^{0.6}$.

\subsection{The case $ m \neq 0$ and $\mu = 0$}
In this case eq.(\ref{11}) yields
\begin{equation}
B (x) \,=\,m + \frac{e^2}{2 \pi} \int dx \, \frac{1}{\mid
x - y \mid} \,\frac{B(y)}{y + B^2(y) }\;.
\label{17}
\end{equation}
and we see that the same argument which proves the non-existence
of non-trivial solutions for the case 1), applies also here.
However, the difference is that now, not even $B = 0$ (or any constant)
is a solution of (\ref{17}).
Because of the infrared divergence in the photon propagator, the ladder
series is ill-defined, and thus provides no consistent solution.
In an improved approximation, one should take into account the fact that the
photon propagator is modified by the fermion-loop correction.
This modification will be particularly important in this case, because
it will tame the infrared divergences by giving the photon a mass.
We conclude from this example that neglecting
the corrections to the photon propagator is not realistic in this case.

\subsection{The case $m \neq 0$ and $\mu \neq 0$}
This case corresponds to the so-called Massive Schwinger Model, which
has many interesting properties~\cite{col}, and is not exactly
solvable. From our point of view, we want to know which is the effect
of an explicit bare fermion mass on the self-consistent solution
of eq.(\ref{12}).

Now we deal with eq.(\ref{12}) as it stands. From a
discussion similar to the one of case 2), we get the constraint
\begin{equation}
m \leq b(0) \leq m + \frac{e^2}{2 \mu^2} \;.
\label{18}
\end{equation}
In Figure 3 we show the behaviour of the numerical solution
of eq.(\ref{12}) for $g = 1$. For fixed $m$, the profile of $b$
is similar to the one of case 2), except for a vertical shift in $m$.
In Figure 4 we show a plot of $b(0)$ as a function of the bare fermion mass.
Note the it is very approximately linear with $m$.
Let us consider this point with more detail.
The explicit mass appears just as an additive constant in the r.h.s in
eq.(\ref{12}).  Then, in the first iteration, which is equivalent to a 
one-loop
calculation, $m$ just shifts the  result corresponding to $m = 0$ by a 
constant
$m/\mu$.  Repeating the iteration, $m$ appears  non-linearly
so it would be naive to expect the solution for $m \neq 0$ to be just
the one for $m = 0$ shifted by a constant.
But the numerical evidence shows that this is the case when $m/\mu$ is
larger than one. The reason is that when $m$ is large compared to $e$, we
can regard the ladder series as an expansion in powers of $e / m$,
the leading term being the one-loop result. Thus we expect the non-linear
part to be suppressed. To get an idea of when we can regard $m$ as
`large', we estimate the relative ratio between the explicit and the 
dynamical
masses after one iteration. $b(0)$ can then be evaluated analytically:
\begin{equation}
b(0) \,=\, \frac{m}{\mu} + \frac{e^2}{\mu^2} f(c) \;,
\label{19}
\end{equation}
where
\begin{equation}
f(c) \,=\, \frac{c}{c^2-1} {\rm ln} c \;,
\label{20}
\end{equation}
and $c$ is the initial value of $b$.  The function $f(c)$ has a
maximum value $f = \frac{1}{2}$ at $c=1$.  Thus, compared with the explicit
term $\frac{m}{\mu}$, the dynamical term gives a contribution less than
$\pi g$ for any initial value of $b$.
Then, when $\frac{m}{\mu} \gg \pi g$, we expect a behaviour
approximately linear with $m$.

Let us see now if in this limiting case the ladder approximation is
reliable.
We will approximate the exact SDE for very large  $m$.
Our approach is to replace $G'$ and $\Gamma$
in (\ref{6}) by their large-$m$ approximations. By large-$m$ we
mean that the dimensionless parameter $e/m$ is very small, as well
as the ratios $p/m$ in the 1PI functions.
Let us consider first the correction to the photon propagator.
Considering $e/m <<1$, it is easy to see that the leading
correction is the usual fermion loop. For very large $m$, the
contribution of this diagram is just
\begin{equation}
\Pi_{\mu \nu} \,=\, - \frac{e^2}{2 \pi} \, g_{\mu \nu}
\label{b}
\end{equation}
where we have also assumed that $m$ is large compared with the external
momenta. Then the full propagator corrected by this function will
be just like the free one, but with $\mu^2$ replaced by
$\mu^2 + e^2 /{\pi}$.

Now, regarding the vertex function, we easily see that the first
correction is of order 3 in $e/m$, and so we neglect it in the
leading approximation, replacing $\Gamma$ by the zeroth order
function, i.e., $\Gamma_{\mu} \simeq \gamma_{\mu}$. Thus, putting
together the approximations for $G'$ and $\Gamma$, the result is
that we get the ladder equation, but with a shifted photon mass.
Now we can argue that the ladder approximation will be
reliable  if $e/m <<1$ and  $e/\mu << 1$. The first condition is to justify 
the
expansion, and the second to neglect the dynamical photon mass
(\ref{b}).

Note that the assumption $k/m << 1$ in the computation of the fermion
loop does not produce any harm at large $k$, because in
that regime the behaviour of the vector field propagator
is dominated by the $k^2$ in the denominator.

Thus, we can also argue that under these conditions, the
usual perturbation
expansion should be reliable, i.e., the main contribution to the
fermion self energy should be given by the one-loop diagram, or
one-iteration with the bare mass as initial self-energy. The
self-consistent calculation will improve this result, giving
a contribution which will be small, but non-analytic in the coupling 
constant.
In Fig. 5 we show a plot of the one-iteration calculation, against
the self-consistent solution for a case where
$m = e/\sqrt{2 \pi}$. Already the
plots agree within a 5 percent error.

We conclude this papers suggesting that the analysis
we have made of this model could be extended to more realistic
ones, where one might be interested in knowing if the non-perturbative
effects are negligible or not in the computation of the
full (bare + dynamical) fermion mass. Note that this aim is independent of
the usual spontaneous chiral symmetry breaking study, because of the
explicit breaking due to the bare fermion mass.

\section*{Appendix: The exact and quenched solutions}
The exact solution of the Schwinger model has been known since
the pionering work by Schwinger~\cite{sch}.
We present here a calculation of the exact fermion propagator
when the gauge field  in the Schwinger Model is massive;
this is a straightforward generalization of the
massless case, but we show it here for the sake of
completeness. It is also easy to discuss the effect of making the
quenched approximation in this framework.
In the path integral framework, the complete fermion propagator is
defined by:
\begin{equation}
S(x-y) \,=\,\langle \psi (x) {\bar \psi} (y) \rangle \,=\,
\frac{1}{N} \, \int {\cal D}A \, {\cal D} \psi \, {\cal D}{\bar \psi} \,\,
\psi (x) {\bar \psi} (y)
\exp (i \int d^2 x {\cal L} )
\label{21}
\end{equation}
where ${\cal L}$ is the Lagrangian of eq.({\ref{1}), and $N$ is a
normalization factor. The integration over the fermionic fields yields:
\begin{eqnarray}
S(x-y) \,&=&\, \int {\cal D} A \,\, \det ( \not \!\! D + i m) \,
\langle x \mid {(\not \!\! D + i m)}^{-1} \mid y \rangle \, \nonumber\\
&\times& \exp [ i \int d^2 x ( - \frac{1}{4} F^2 + \frac{1}{2} \mu^2 A^2 -
\frac{\lambda}{2} {(\partial \cdot A)}^2 ) ] \;.
\label{22}
\end{eqnarray}
Both the determinant and the inverse of the Dirac operator $\not \!\!D
= \not \! \partial + i e \not \!\! A $  can be calculated exactly:
\begin{equation}
\det ( \not \!\! D + i m ) \,=\, \exp ( - i \int d^2 x
\frac{e^2}{4 \pi}
F^{\mu \nu} \frac{1}{\partial^2} F_{\mu \nu} )  \;,
\label{23}
\end{equation}
\begin{eqnarray}
S (x-y) \,&=&\, \exp \{ - i e  [ {\partial}^{-2} \partial \cdot A (x) -
e {\partial}^{-2} \partial \cdot A(y)  \nonumber\\
&+& \gamma_5 ( {\partial}^{-2} \epsilon^{\mu \nu} \partial_{\mu} A_{\nu} (x) 
-
{\partial}^{-2} \epsilon^{\mu \nu} \partial_{\mu} A_{\nu} (y) ] \} \;.
\label{24}
\end{eqnarray}
Inserting  the results eqs.(\ref{23}) and (\ref{24}) into eq.(\ref{22}), and
 performing the (Gaussian)
integration over $A$, we obtain (now in Euclidean spacetime):
\begin{equation}
S(x-y) \,=\, S_0 (x-y) \,
e^{ K_1 (0) - K_1 (x-y) } \, e^{ - K_2 (0) + K_2 (x-y) } \;,
\label{a25}
\end{equation}
with the definitions:
\begin{eqnarray}
K_1 (x) \,&=&\, e^2 \int \frac{d^2 k}{ {(2 \pi)}^2 }
\frac{1}{k^2
(k^2 + \mu^2 + e^2 /\pi)} e^{i k \cdot x} \nonumber \\
K_2 (x) \,&=&\, \frac{e^2}{\lambda} \int \frac{d^2 k}{ {(2 \pi)}^2 }
\frac{1}{k^2
(\lambda k^2 + \mu^2)} e^{i k \cdot x} \;.
\label{26}
\end{eqnarray}
>From eq.(\ref{a25}), one easily sees that the fermion remains massless
since there is no extra pole at $p^2 \neq 0$.

It is very easy in this approach to understand what happens in the
`quenched' approximation. This is tantamount of discarding the
determinant in eq.(\ref{23}), replacing it by 1. In diagrammatic language, it
is equivalent to summing all the diagrams which contain one electron
line and any possible correction due to photon lines. Equivalently,
one can add to the rainbow diagrams the ones where the photons'
propagators cross.
Note that this approximation is gauge-invariant, what is trivial
from the fact that we neglect a gauge-invariant quantity. This
is achieved in the diagrammatic language by the crossed diagrams,
which correct the vertex in such a way that the Ward identity is
satisfied.

In this approximation, we get also a massless fermion, as can
be seen from eq.(\ref{a25}), disregarding the determinant's contribution.
This leads to
\begin{equation}
S(x-y)\,=\,S_0(x-y) \,
\end{equation}
in the gauge $\lambda = 1$.

\section*{Acknowledgements}
C. D. F. was supported by an European Community Postdoctoral Fellowship.
T. M. was supported in part by the British Council and the Daiwa
Anglo-Japanese Foundation. We also would like to express our
acknowledgement to Dr. I. J. R. Aitchison for his kind hospitality.

\newpage
{\bf Figure captions}

\vspace{0.3 cm}

{\bf Figure 1} The self-energy $b(s)$  for five different values of
$g = \frac{e^2}{2 \pi \mu^2}$ ( case 2 ).

\vspace{0.3 cm}

{\bf Figure 2} The zero-momentum self-energy $b(0)$ for some values of $g$,
fitted by $b(0) = 1.29 \, g^{0.6}$ ( case 2 ).

\vspace{0.3 cm}

{\bf Figure 3} The self-energy $b(s)$ for $g=1$ and $m = \mu$
( case 4 ).

\vspace{0.3 cm}

{\bf Figure 4} The zero-momentum self-energy $b(0)$ for $g = 1$ as
a function of the ratio between the fermion and boson bare masses.
The linear plot is the one-loop result ( case 4 ).

\vspace{0.3 cm}

{\bf Figure 5} The result of iterating the SDE (12) (for
$g = 1$) once, starting from the initial value $b = constant =
\frac{m}{\mu}$ against the self-consistent solution of the same equation
( case 4 ).

\end{document}